\documentstyle[12pt]{article}
\title{A Wilson Renormalization Group\\
Approach to Light-Front\\
Tamm-Dancoff\\
Scalar Field Theory}
\author{Edsel A. Ammons\\
Department of Physics, The Ohio State University\\
174 West 18th Avenue, Columbus, Ohio 43210}
\date{July 24, 1994}
\begin{document}
% The following 3 lines are for preprints only.
%\noindent{\it Accepted for publication in:} \hfill {OSU-NT-94-199}

%\noindent{\it Phys. Rev. D}

%\vspace{12pt}
\baselineskip=18pt
\maketitle
{\Large {\bf Abstract}}

A program to utilize the Tamm-Dancoff approximation,
on the light-front, to solve relativistic quantum field
theories, is presented.  We present a well defined renormalization
program for the Tamm-Dancoff approximation.  This renormalization
program utilizes a Minkowski space version of Wilson's
renormalization group.  We studied light-front $\phi^{4}$
field theory in 3+1 dimensions, within a two-particle truncation
of Fock space.  We further simplified our calculations by
considering only one marginal operator and one irrelevant
operator.  The renormalization procedure required no more
marginal or relevant operators.  We derived an effective,
renormalized, Hamiltonian.  These techniques may be germane
to the effort to find an effective, low energy, light-front
Hamiltonian for quantum chromodynamics.\\
PACS number(s): 11.10Gh, 11.10Ef
\newpage
\section{Introduction}
As a part of a Hamiltonian approach to relativistic quantum
field theory, we report here on a
method which is partly analytical and partly numerical.
It is a method for generating an effective Hamiltonian that has
all the physics of the theory, below a
cut-off energy scale, and
within a certain Tamm-Dancoff truncation of Fock space
\cite{1,2}.  That is, a
given truncation of Fock space will likely describe some
classes of states more accurately than other classes.
This paper describes a careful analysis of the renormalization
problems of a simplified Tamm-Dancoff truncation of the
scalar field.  This simplification, used in order to bring out
the essential features of the method, consists in considering
the approximation of one marginal operator and one irrelevant
operator.
We do not investigate the validity of the Tamm-Dancoff
truncation in this paper, concentrating only on reproducing
the effects of high energy states within the truncation.
We use a renormalization group procedure, used previously
\cite{3}, which eliminates high energy states, generating an
effective Hamiltonian for the low energy states.
That is, the high energy sector of the theory is removed in
a manner that preserves the physics of the low energy
sector.  The low energy physics, that is within a given
Tamm-Dancoff truncation of Fock space, is preserved as the
high energy sector is removed,
as we show below.  What sets the boundary between the high energy
scales and low energy scales are the masses of the bound states
one wishes to study.  The masses of interest should lie in
the low energy scale region.

Elimination of the high energy sector also eliminates divergences.
Hence, regularization of the Hamiltonian is achieved.
Renormalization is achieved because the real physical effects
of the high energy sector upon the low energy sector are represented
by the systematic inclusion of additional interactions.

Scalar field theory is used in order to provide as simple a
framework as possible for displaying the machinery of the
Tamm-Dancoff truncated, light-front field theory with the
Minkowski space version of Wilson's renormalization
group \cite{4}.
For now we simply wish to develop
further a realization of the Wilson conception of renormalization
in field theory
within the context of the light-front Tamm-Dancoff formalism
\cite{5}.

Either an explicit or implicit
Tamm-Dancoff truncation of Fock space is essential, because the
complete vector space of a quantum field theory cannot be
put on a computer.  This anticipates the eventual need for
a numerical solution of a theory.
In contradistinction to Perry's study \cite{3}, a
Tamm-Dancoff truncation is adopted at the very beginning of the
analysis.  One result is that the $\beta$-function for the
truncated scalar theory is not the same as that of the nontruncated
theory.  This does not represent a problem with our approach over
and beyond those presented by the fact of a Tamm-Dancoff truncation
of Fock space.
The parameters will run in a manner that is appropriate for the
given truncation of the theory.
We do not comment on the implications of this
fact for the validity of the Tamm-Dancoff truncation except to
say that the Tamm-Dancoff approximation is a variational
approach to field theory.  The larger the Fock space that is
utilized, the more exactly will the variational solutions correspond
to the exact solutions for the observables of the theory.
These observables, of course, include the
physical particle masses, which are invariant
masses of the energy-momentum eigenstates, and particle form
factors, derived from the particle wavefunctions.  Scattering
amplitudes for scattering between these
particles can also be calculated.  Also, as more of the Fock space
is included, the parameters will run more analogously to the way
the parameters run in the nontruncated theory.

The use of the light-front formalism is desirable, if not essential,
because of the restriction to positive longitudinal momenta
\cite{6}.  This
allows all vacuum structure to be forced into the zero-mode structure
of a theory.  If the vacuum structure can be thus isolated, it may
be possible to replace this structure with interactions.
This procedure is straightforward only in light-front coordinates.
With the light-front formalism, the problem of solving for excited
state wavefunctions, the particles, is separated from the problem
of solving for the ground state wavefunction, the physical vacuum, within the
Tamm-Dancoff approximation.  This is simply because the physical
vacuum structure has been greatly simplified.  The vacuum is
`trivial'.  That is, the physical vacuum is also the no-constituent
state.
This is not the case in equal-time
Tamm-Dancoff.  With this simplified vacuum, however, must come
a complication of the operators of the theory, such as the
Hamiltonian.  In addition, the complicated operator structure
of the formalism may result in a simplified excited state
structure, thus providing ultimate justification of the Tamm-Dancoff
truncation.

The starting point of this approach is a fixed-point
Hamiltonian of a Wilson renormalization group transformation.
There may be more than one fixed-point
for a given canonical theory, and any number of these fixed-points
may or may not be experimentally relevant.  We have
chosen here to work with a Gaussian, massless fixed-point
Hamiltonian, a Hamiltonian consisting only of a kinetic term.
Generation of the canonical Hamiltonian is an essential but
secondary aspect of the method.  The canonical Hamiltonian
contains all the relevant and marginal operators consistent
with the symmetries of the theory.  We assume that the couplings
for these operators are those which should run independently
with the cut-off \cite{7}.
Generation of the effective Hamiltonian is an analytical and
numerical problem.  Solution of the effective Hamiltonian may
then be a non-perturbative, tractable, numerical problem.

Another way is now open for the non-perturbative solution of
relativistic quantum field theories.  The specific realizations
of Wilson renormalization groups must be perturbative
about the fixed-point Hamiltonian at present, and
this may at times be a limitation.  However, the effective Hamiltonian,
in the truncated Fock space, may be open to an exhaustive numerical
solution on present-day computers.
The non-perturbative solution of quantum chromodynamics may be
approachable by this method.

Perry has outlined the Wilson renormalization group approach
to light-front field theory \cite{3}.  This report adds a
Tamm-Dancoff truncation to obtain a simple, testable example
of the approach and
suggests the plausibility of the whole approach.
An understanding of the Wilson renormalization group in the
detail presented in Wilson and Kogut \cite{4} or in
Wilson \cite{8} is desirable but not essential.
Familiarity with Perry's study, which inaugurated this approach
\cite{3}, is also desirable.

The next section describes the calculation of a Bloch transformation
and its effects upon an observable, the two-particle to two-particle
scattering amplitude within a two-particle truncation of Fock
space.
Section 3 discusses these calculations and
presents the resulting effective Hamiltonian.  Section 4 presents
an outline for broader calculations, mentioning some important
remaining issues which are a part of the generation of effective
Hamiltonians for any field theory within
light-front Tamm-Dancoff, and concludes
with a summary.
\section{Calculations}
\subsection{A Wilson Renormalization Group Approach}
The Wilson renormalization group of Perry \cite{3} is used
in this report.
The starting point is an unstable, ultraviolet
fixed-point Hamiltonian.
The
basic ingredients of a Wilson renormalization group transformation
are:
\begin{enumerate}
\item A Bloch transformation is the
Minkowski space counterpart of a Kadanoff
transformation in Euclidean space
\cite{3,9,10,11,12}
which involves elimination of
energy scales, progressing from ultraviolet to infrared.
That is, the cut-off is lowered.  Bloch
transformations must typically be realized
perturbatively, because presently this is the only
technique generally available.
This is reasonable for small enough values of the
couplings.
\item Rescaling of the remaining energy scales
and rescaling of the field variables must be carried out.
This must be done so that a fixed-point
Hamiltonian, which is an invariant of the complete transformation,
may exit.
\end{enumerate}
An explanation of the interrelationship of the canonical approach
to field theory and the Wilson fixed-point approach is given
in Wilson and Kogut \cite{4}.

Although there may be a number of different kinds of topological
properties of the allowed space of Hamiltonians which will
enable the procedure of cut-off lowering and rescaling to
generate renormalized Hamiltonians, the following is a possible
scenario;
there are at least two fixed-point Hamiltonians for the given
Wilson renormalization group transformation.
The infrared one must be stable.
All trajectories in the neighborhood of the fixed-point must flow
into the fixed-point.
The ultraviolet one must be unstable. This means not all of the trajectories
in the neighborhood of the fixed-point flow into the fixed-point.
This scenario is likely true for the scalar field.  The infrared
fixed-point Hamiltonian will consist of a kinetic term and a mass
term.

Essentially, the cut-off must eliminate eigenstates of ${\rm H}^{*}$,
the ultraviolet fixed-point Hamiltonian.
Since one is eliminating energy levels of ${\rm H}^{*}$, in order
to follow the trajectory of Hamiltonians, one wants
to rescale energy, in general.  At the end of the calculation, in
deriving the effective Hamiltonian, the rescalings are undone, and
the original scale is restored.
${\rm H}^{*}$ will be chosen to be a Gaussian, no interactions,
massless Hamiltonian.
Fig. 1 is a restatement of Fig. 12.7 of Wilson and Kogut
\cite{4},
presented here for purposes of continuity, which is an
expression of the above topological scenario.  ${\rm H_{D}}$ is a
surface of cut-off canonical Hamiltonians, with cut-off
$\Lambda_{0}$.  As $\Lambda_{0}$ is made to go toward infinity,
the parameters of the Hamiltonian on ${\rm H_{D}}$ are imagined
to vary with this cut-off such that the canonical surface
will intersect the critical surface C of a fixed-point Hamiltonian,
here called $P_{\infty}$, at infinite $\Lambda_{0}$.
The important points are that Hamiltonians, in this space,
residing on the critical surface, are driven into $P_{\infty}$
by a large number of renormalization group transformations.
All other Hamiltonians in the space are driven into another
fixed-point, called $P_{0}$, by a large number of transformations.

A Bloch transformation lowers the cutoff by a finite amount,
and the final cutoff, $\Lambda_{f}$, is given by
$$\Lambda_{f}=\Lambda_{0}\cdot \frac{\Lambda_{1}}{\Lambda_{0}}
\cdot \frac{\Lambda_{1}'}{\Lambda_{0}}\cdot \frac{\Lambda_{1}''}
{\Lambda_{0}}....,$$
where each $\Lambda_{1}$, in each factor, is associated with
a given application of a Bloch transformation.  That is, after
a complete renormalization group transformation, the resultant
Hamiltonian has the same cutoff as the original Hamiltonian, which
is $\Lambda_{0}$, because of the rescalings,
but from the point of view of the world
described by the starting Hamiltonian, each successive application
of the renormalization group transformation eliminates lower
and lower energy scales.
If $\Lambda_{0}$ is infinite, $\Lambda_{f}$ will be infinite,
and if $\Lambda_{0}$ is finite, $\Lambda_{f}$ will be zero
after an infinite number of transformations.  So the $\Lambda_{f}$
associated with $P_{\infty}$ is infinite and the $\Lambda_{f}$
associated with $P_{0}$ is zero.  Hence, the former fixed-point
is called `ultraviolet', and the latter is called an `infrared'
fixed-point.

Starting with a Hamiltonian on the canonical surface and
applying renormalization group transformations, a trajectory
will emanate, eventually going into the infrared fixed-point,
if $\Lambda_{0}$ is finite.  As one makes $\Lambda_{0}$
larger and larger, the trajectory will hug the critical surface,
more and more, approaching the ultraviolet fixed-point, before
eventually diverging from it, going onward toward $P_{0}$.
The consequence of this is that the Hamiltonians Q, at the
cutoff 1 GeV, for example, approach a limiting Hamiltonian,
$Q_{\infty}$.  $Q_{\infty}$ describes all the physics of
the uncut-off field theory, but does so only below the cutoff
of 1 GeV.  This is the renormalized Hamiltonian at the cutoff
$\Lambda_{f}=1$ GeV.  Simply cutting the canonical Hamiltonian
off at 1 GeV is a zeroth-order approximation to the renormalized
Hamiltonian.  That is $Q_{1}$.  $Q_{2}$ is a better approximation.
$Q_{5}$ is even better, and so forth.

In our work below,
we choose a Hamiltonian that is at a point on a trajectory
that is near the ultraviolet fixed-point, $P_{\infty}$.  The
Hamiltonian will consist of the ultraviolet fixed-point Hamiltonian
plus a perturbing interaction.    For the Gaussian, interactionless,
massless ultraviolet fixed-point used below,
one can usually begin with a
perturbation consisting of all the relevant and marginal operators
of the canonical Hamiltonian of the theory.  One then determines
all the additional relevant and marginal operators generated by
applications of the renormalization group transformation, within
the given Fock space truncation.  Irrelevant operators will also
be generated.  The new perturbation will then consist of all
these relevant and marginal operators and usually of only
a subset of the irrelevant operators.  The terms `relevant',
`marginal', and `irrelevant' are further explained below.
\subsection{Bloch transformation}
To illustrate the techniques of the renormalization group approach
to field theory, a Gaussian, massless, ultraviolet
fixed-point Hamiltonian
will be assumed \cite{3,5}.  The renormalization
group transformation of Perry \cite{3} will be applied here.
For this transformation, a dimensionless parameter, coupling,
multiplying a given operator in the Hamiltonian, is classified
as relevant, marginal, or irrelevant according to the following
criteria:

For the linearized renormalization group transformation,
which, for this case, is simply the rescaling part of
the complete transformation, the operator
will satisfy the following eigenvalue equation,
\begin{eqnarray}
L\cdot O&=&\rho O\;,
\end{eqnarray}
where $L$ is the linearized transformation.  We can label $O$ with
a subscript that displays the number of field operators in $O$
and a superscript that displays the number of powers of transverse
momenta.  For example (see Appendix for notational definitions),
\begin{eqnarray}
O^{6}_{2}&=&\int d\tilde{q}\frac{\vec{q}^{\:4}}{q^{+}}|q><q|\;,
\end{eqnarray}
where Eqns. (36) and (41) in the appendix enable one
to see that the projection operator in Eqn. (2) is made
from a product of two field operators with the vacuum projector.

Applying $L$ to any operator $O^{m}_{n}$ one finds,
\begin{eqnarray}
L\cdot O^{m}_{n}&=&(\frac{\Lambda_{1}}{\Lambda_{0}})^{(m-n-2)}
O^{m}_{n}\;,
\end{eqnarray}
where $\Lambda_{0}$ is an initial cutoff on the invariant
mass of allowed states, and $\Lambda_{1}$ is the cutoff
after a Bloch transformation.
Relevant couplings multiply operators for which $\rho >1$.
Since $\Lambda_{0}>\Lambda_{1}$,
the meaning is now clear.  Marginal couplings
multiply operators for which $\rho =1$.  Irrelevant couplings
correspond to $\rho <1$.  The operators are classified likewise.

When ${\rm H^{*}}$ is chosen to consist of only a kinetic term,
$L$ is easily constructed.  Let $T$ be the full renormalization
group transformation.  Let ${\rm H}_{l}$ be the Hamiltonian
resulting from $l$ operations of the renormalization group
transformation on the ultraviolet fixed-point plus its perturbation.
That is,
\begin{eqnarray}
{\rm H}_{l}&=&{\rm H}^{*}+\delta {\rm H}_{l} \nonumber \\
{\rm H}^{*}+\delta {\rm H}_{l+1}&=&T[{\rm H}^{*}+\delta {\rm H}_{l}]
\nonumber \\
&=&{\rm H}^{*}+
L\cdot \delta{\rm H}_{l}+{\cal O}(\delta {\rm H}^{2}_{l})
\end{eqnarray}
These equations define the linear operator $L$.  The form of the
Bloch transformation is given, to second-order, below.  Then it
follows from this form, and from Eqn. (4) above that the linearized
$T$, which is $L$, is just the rescaling operation.  The rescalings
are, in turn, designed to leave ${\rm H}^{*}$ invariant.  Eqn. (3)
is the resulting scaling relation.

See Eqns. 3.3, 3.6, and 3.7 of Perry \cite{3} for the general
scalar Hamiltonian allowed, by power counting, in our space of Hamiltonians.
The functions $u^{(m,n)}_{j}$ below are the generalized interactions
associated with the $\phi^{j}$ interaction that connects the $m$-particle
sector with the $n$-particle sector.
We truncate Fock space to allow only the one and two-particle
sectors, where the general Hamiltonian is,
\begin{eqnarray}
{\rm H}&=&\int d\tilde{q}\frac{u^{(1,1)}_{2}(q)}{q^{+}}|q><q| \nonumber\\
 &+&\int d\tilde{q}_{1}d\tilde{q}_{2}(\frac{u_{2}^{(2,2)}(q_{1})}
{q_{1}^{+}}+\frac{u_{2}^{(2,2)}(q_{2})}{q_{2}^{+}})
|q_{1},q_{2}><q_{1},q_{2}| \nonumber\\
 &+&\frac{1}{4}\int d\tilde{q}_{1}d\tilde{q}_{2}d\tilde{q}_{3}
\frac{u_{4}^{(2,2)}(q_{1},q_{2},-q_{3},-q_{1}-q_{2}+q_{3})}
{q_{1}^{+}+q_{2}^{+}-q_{3}^{+}}|q_{1},q_{2}><q_{3},q_{1}+q_{2}-q_{3}|
\: .
\end{eqnarray}

The perturbation of the fixed-point will, at first, be assumed
to be a marginal operator in the $\phi^{4}$ interaction.  That is,
initially $u_{4}=\lambda$, a constant (see Appendix).
There are no marginal or relevant
operators in the $\phi^{6}$ interaction, for example.  There are
relevant and marginal operators in the $\phi^{2}$ interaction.
There are irrelevant operators in all interactions.
As a result,
\begin{eqnarray}
{\rm H}_{\epsilon_{0}}&=&h+v\; ,
\end{eqnarray}
at an initial cut-off of $\epsilon_{0}$.  This cut-off
corresponds to an
invariant mass of $\Lambda_{0}$ where,
\begin{eqnarray}
\epsilon_{0}&=&\frac{\vec{P}^{\:2}+
\Lambda_{0}^{2}}{P^{+}} \; .\nonumber
\end{eqnarray}

The ultraviolet fixed-point Hamiltonian, $h={\rm H}^{*}$, is
\begin{eqnarray}
h&=&\int d\tilde{q}_{1}d\tilde{q}_{2}\theta (\frac{\vec{P}^{\:2}
+\Lambda_{0}^{2}}{P^{+}}-\frac{\vec{q}_{1}^{\:2}}{q_{1}^{+}}-
\frac{\vec{q}_{2}^{\:2}}{q_{2}^{+}})(\frac{\vec{q}_{1}^{\:2}}{q_{1}^{+}}
+\frac{\vec{q}_{2}^{\:2}}{q_{2}^{+}})|q_{1},q_{2}><q_{1},q_{2}| \; ,
\end{eqnarray}
where $\vec{P}=\vec{q}_{1}+\vec{q}_{2}$ and $P^{+}=q_{1}^{+}+q_{2}^{+}$.
The $\theta$-function displays the cut-off.
With this fixed-point, the bosons of the basis set are massless.
The perturbation, $v$, is given by,
{\samepage
\begin{eqnarray}
v&=&\frac{\lambda}{4}\int d\tilde{q}_{1}d\tilde{q}_{2}d\tilde{q}_{3}
\frac{|q_{1},q_{2}><q_{3},q_{1}+q_{2}-q_{3}|}{q_{1}^{+}+q_{2}^{+}
-q_{3}^{+}}\\
 & &\theta(q_{1}^{+}+q_{2}^{+}-q_{3}^{+})\theta(\frac{\vec{P}^{\:2}
+\Lambda_{0}^{2}}{P^{+}}-\frac{\vec{q}_{1}^{\:2}}{q_{1}^{+}}-
\frac{\vec{q}_{2}^{\:2}}{q_{2}^{+}})\theta(\frac{\vec{P}^{\:2}+
\Lambda_{0}^{2}}{P^{+}}-\frac{\vec{q}_{3}^{\:2}}{q_{3}^{+}}-
\frac{(\vec{q}_{1}+\vec{q}_{2}-\vec{q}_{3})^{2}}{q_{1}^{+}+q_{2}^{+}-
q_{3}^{+}}) \: .\nonumber
\end{eqnarray}}

The Bloch transformation
lowers the cut-off by eliminating more
eigenstates of ${\rm H}^{*}$.  The resulting Hamiltonian, ${\rm H}_{
\epsilon_{1}}$, has the same low-lying eigenvalues
as those of ${\rm H}_{\epsilon_{0}}$.
The matrix elements of ${\rm H}_{\epsilon_{1}}$, to second order
in $v$, are \cite{3},
\begin{eqnarray}
<a|{\rm H}_{\epsilon_{1}}|b>&=&<a|h+v|b> \nonumber \\
 &+& \frac{1}{2}\sum_{i}(\frac{<a|v|i><i|v|b>}{\epsilon_{a}-
\epsilon_{i}}+\frac{<a|v|i><i|v|b>}{\epsilon_{b}-\epsilon_{i}})\: ,
\end{eqnarray}
where $|a>$, $|b>$ are eigenstates of $h$ with eigenvalues
$\epsilon_{a}$ and $\epsilon_{b}$ respectively.  These eigenvalues
are below the new cut-off, $\epsilon_{1}$.
The states $|i>$ are eigenstates of $h$ to be eliminated, with
eigenvalues $\epsilon_{i}$.
$\epsilon_{1}\leq \epsilon_{i}\leq \epsilon_{0}$
($\Lambda_{1}\leq M_{i}\leq \Lambda_{0}$).  Also,
\begin{eqnarray}
\epsilon_{a}&=&\frac{\vec{q}_{1a}^{\:2}}{q_{1a}^{+}}+
\frac{\vec{q}_{2a}^{\:2}}{q_{2a}^{+}}=
\frac{\vec{P}^{\:2}+
M_{a}^{2}}{P^{+}} \; ,\nonumber \\
\epsilon_{i}&=&\frac{\vec{q}_{1i}^{\:2}}{q_{1i}^{+}}+
\frac{\vec{q}_{2i}^{\:2}}{q_{2i}^{+}}=
\frac{\vec{P}^{\:2}+
M_{i}^{2}}{P^{+}} \; ,\nonumber \\
\epsilon_{b}&=&\frac{\vec{q}_{1b}^{\:2}}{q_{1b}^{+}}+
\frac{\vec{q}_{2b}^{\:2}}{q_{2b}^{+}}=
\frac{\vec{P}^{\:2}+
M_{b}^{2}}{P^{+}} \; .
\end{eqnarray}
Now the first second-order term from Eqn. (9) is,
\begin{eqnarray}
\frac{1}{2}\sum_{i}\frac{<a|v|i><i|v|b>}{\epsilon_{a}-\epsilon_{i}}
&=& <a|v'|b>
\end{eqnarray}
where $v'$ is $v$ with $u_{4}$ replaced by $\delta u_{4}$, and
\begin{eqnarray}
\delta u_{4}&=&-\frac{\lambda^{2}}{64\pi^{2}}\int_{0}^{1}dx
\int_{\Lambda_{1}^{2}x(1-x)}^{\Lambda_{0}^{2}x(1-x)}
\frac{dr^{2}}{\vec{r}^{\:2}}\sum_{n=0}^{\infty}(\frac{x(1-x)}
{z(1-z)}\frac{\vec{s}^{\:2}}{\vec{r}^{\:2}})^{n}\; .
\end{eqnarray}
$\Lambda_{1}$ is the invariant mass cut-off associated with
$\epsilon_{1}$, and
\begin{eqnarray}
\frac{\vec{s}^{\:2}}{z(1-z)}& <&  \Lambda_{1}^{2}
  \leq  \frac{\vec{r}^{\:2}}{x(1-x)}  \leq  \Lambda_{0}^{2}
 \; ; \nonumber \\
\end{eqnarray}
where,
\begin{eqnarray}
\epsilon_{a}&=&\frac{\vec{P}^{\:2}}{P^{+}}+\frac{\vec{s}^{\:2}}
{P^{+}z(1-z)} \; ,\\
\epsilon_{i}&=&\frac{\vec{P}^{\:2}}{P^{+}}+\frac{\vec{r}^{\:2}}
{P^{+}x(1-x)}\; ,
\end{eqnarray}
and,
\begin{eqnarray}
q_{1a}&=&(zP^{+},z\vec{P}+\vec{s}) \; ,\nonumber \\
q_{2a}&=&((1-z)P^{+},(1-z)\vec{P}-\vec{s})\; .
\end{eqnarray}
$(z,\vec{s}\:)$ are Jacobi coordinates for $|a>$,
$(x,\vec{r}\:)$ are Jacobi coordinates for $|i>$, and we
now introduce $(y,\vec{t}\:)$ as Jacobi coordinates for $|b>$.

$\lambda$ multiplies a marginal operator contribution to $u_{4}$.
$\delta u_{4}$ consists of a marginal contribution,
$n=0$ in Eqn. (12), and an
infinite number of irrelevant operator contributions,
$n\geq 1$ in (12).  Considering
the term with $\epsilon_{b}$ in the denominator as well,
the other second-order term in Eqn. (9), and
ignoring all the irrelevant operator contributions,
\begin{eqnarray}
\delta u_{4}&=& -\frac{\lambda^{2}}{32\pi^{2}}
\int_{0}^{1}\int_{\Lambda_{1}^{2}x(1-x)}^{\Lambda_{0}^{2}x(1-x)}
\frac{dr^{2}dx}{\vec{r}^{\:2}} \nonumber \\
&+&{\cal O}(\vec{s}^{\:2})+{\cal O}(\vec{t}^{\:2}) \nonumber \\
 &=& -\frac{\lambda^{2}}{16\pi^{2}}ln(\frac{\Lambda_{0}}{\Lambda_{1}})
 +{\cal O}(\vec{s}^{\:2})+{\cal O}(\vec{t}^{\:2})\;.
\end{eqnarray}
Keeping the leading irrelevant operator contributions to $u_{4}$,
$n=1$ in Eqn. (12),
which are generated by a Bloch transformation, at second order,
\begin{eqnarray}
\delta u_{4} &=& -\frac{\lambda^{2}}{16\pi^{2}}ln(\frac{\Lambda_{0}}
{\Lambda_{1}})-\frac{\lambda^{2}}{64\pi^{2}}(\frac{1}{\Lambda_{1}^{2}}
-\frac{1}{\Lambda_{0}^{2}})\frac{\vec{s}^{\:2}}{z(1-z)}
+{\cal O}(\vec{s}^{\:4}) \nonumber \\
& &-\frac{\lambda^{2}}{64\pi^{2}}(\frac{1}{\Lambda_{1}^{2}}
-\frac{1}{\Lambda_{0}^{2}})\frac{\vec{t}^{\:2}}{y(1-y)}
+{\cal O}(\vec{t}^{\:4})+{\cal O}(\vec{s}^{\:2}\vec{t}^{\:2})\;.
\end{eqnarray}

Let $u_{4}$, in $v$, consist only of contributions from the leading
irrelevant operator. This contribution is $\frac{\alpha}{\Lambda_{0}^{2}}
(\frac{\vec{s'}^{\:2}}{z'(1-z')}+
\frac{\vec{t'}^{\:2}}{y'(1-y')})$, where $\alpha$ is the irrelevant
coupling strength, and $(z',\vec{s'})$ is made from $q_{1}$ and $q_{2}$
in $v$, and $(y',\vec{t'})$ is made from $(q_{3},q_{1}+q_{2}-q_{3})$
in $v$.
The second order Bloch transformation
gives, to the leading irrelevant operator,
\begin{eqnarray}
\delta u_{4}&=& -\frac{3\alpha^{2}}{64\pi^{2}}
(1-\frac{\Lambda_{1}^{2}}{\Lambda_{0}^{2}})\frac{1}{\Lambda_{0}^{2}}
(\frac{\vec{s'}^{\:2}}{z'(1-z')}+\frac{\vec{t'}^{\:2}}{y'(1-y')})
\nonumber \\
& &-\frac{\alpha^{2}}{64\pi^{2}}(1-\frac{\Lambda_{1}^{4}}{\Lambda_{0}^{4}}
)+{\cal O}(\vec{s'}^{\:4})+{\cal O}(\vec{t'}^{\:4})
+{\cal O}(\vec{s'}^{\:2}\vec{t'}^{\:2})\;.
\end{eqnarray}
These results correspond to the diagrams
in Figs. 3a and 3d, respectively, keeping
the marginal and leading irrelevant operators, where the incoming
and outgoing lines are the states $|a>$, $|b>$, and the vertices
are the interaction $v$, and the internal lines are the states $|i>$.

Fig. 2a displays the vertex  associated with the marginal
coupling $\lambda$.
Fig. 2b shows the leading irrelevant vertex, where the leading
irrelevant operator with strength
$\alpha$ is associated
with the vertex.
Figs. 3b and 3c must also be included in a second order analysis.
The four second-order diagrams, 3a, 3b, 3c, and 3d,
and the two first order diagrams, 2a and 2b,
illustrate how the Bloch transformation can be summarized
as a change of coupling strengths.  That is, these parameters run.

The difference equations that represent the leading effects of the Bloch
transformation are, using $4!g=\lambda$ and $4!w=\alpha$,
\begin{eqnarray}
g'_{n+1}&=&g_{n}-\frac{3g_{n}^{2}}{2\pi^{2}}ln\frac{\Lambda_{0}}
{\Lambda_{1}}-\frac{3g_{n}w_{n}}{2\pi^{2}}(1-\frac{\Lambda_{1}
^{2}}{\Lambda_{0}^{2}})  ,\nonumber \\
& &-\frac{3w_{n}^{2}}{8\pi^{2}}(1-\frac{\Lambda_{1}^{4}}{
\Lambda_{0}^{4}}) \nonumber \; ,\\
w'_{n+1}&=&w_{n}-\frac{9w_{n}^{2}}{8\pi^{2}}(1-\frac{\Lambda_{1}
^{2}}{\Lambda_{0}^{2}})-\frac{3g_{n}w_{n}}{\pi^{2}}ln\frac{\Lambda_
{0}}{\Lambda_{1}} \nonumber \\
 & & -\frac{3g_{n}^{2}}{8\pi^{2}}(\frac{\Lambda_{0}^{2}}{\Lambda_
{1}^{2}}-1) \; ,
\end{eqnarray}
where the parameter set ($g'_{n+1}$,$w'_{n+1}$) is generated from the
parameter set ($g_{n}$,$w_{n}$) by a single Bloch transformation.
A parameter set ($g_{n+1}$,$w_{n+1}$) results from
($g_{n}$,$w_{n}$) by a complete renormalization group transformation.
This transformation is given below.

\subsection{T-matrix}
We now calculate an observable.  If the above approximation to
the full Bloch transformation is
reasonable, this observable should be nearly independent of the cut-off,
below the new cut-off.  We use the
two-particle to two-particle scattering amplitude, for interacting
but massless scalar bosons, to test the procedure.

The S-matrix is given in terms of the T-matrix by \cite{13},
\begin{eqnarray}
S&=&\delta_{ij}-2\pi i\delta(E_{i}-E_{j})<\phi_{i}|T(E_{i})|
\phi_{j}> \; ,
\end{eqnarray}
where $|\phi_{i}>$, $|\phi_{j}>$ are two-particle eigenstates
of $h$ in Eqn. (7).
The T-matrix is given by the perturbative series,
\begin{eqnarray}
T(E)&=&<\phi_{i}|{\rm H}_{{\rm I}}|\phi_{j}>+<\phi_{i}|
{\rm H}_{{\rm I}}\frac{1}{E-{\rm H}_{0}+i\epsilon}{\rm H}_
{{\rm I}}|\phi_{j}>\:+... \; .
\end{eqnarray}
Now, assuming ${\rm H}_{{\rm I}}$ consists only of the generalized
$\phi^{4}$ interaction, that is,
\begin{eqnarray}
u_{4}^{(2,2)}=4!(g+\frac{w\vec{s'}^{\:2}}{\Lambda_{0}^{2}z'(1-z')}
+\frac{w\vec{t'}^{\:2}}{\Lambda_{0}^{2}y'(1-y')})\; ,
\end{eqnarray}
we have the following results,
{\samepage
\begin{eqnarray}
<\phi_{i}|{\rm H}_{{\rm I}}|\phi_{j}>&=&384\pi^{3}(g+w
\frac{2M^{2}}{\Lambda_{0}^{2}})\delta (p_{1i}+p_{2i}-p_{1j}-p_{2j})
 \: ,\\
<\phi_{i}|{\rm H_{I}}\frac{1}{E_{i}-{\rm H}_{0}
+i\epsilon}{\rm H}_{{\rm I}}
|\phi_{j}>&=& -\delta (p_{1i}+p_{2i}-p_{1j}-p_{2j})\cdot
 \nonumber \\
\lefteqn{288\pi [(g+\frac{2wM^{2}}{\Lambda_{0}^{2}})^{2}
ln(\frac{\Lambda^{2}}{M^{2}}-1)
+2w(g+\frac{2wM^{2}}{\Lambda_{0}^{2}})\frac{\Lambda^{2}}
{\Lambda_{0}^{2}}} \nonumber \\
\lefteqn{+\frac{w^{2}}{\Lambda_{0}^{4}}
(\frac{\Lambda^{4}}{2}-M^{2}\Lambda^{2})
+i\pi (g+\frac{2wM^{2}}{\Lambda_{0}^{2}})^{2}]\; ,}
\end{eqnarray}}
where $E_{i}=E_{j}$,
$M$ is the invariant mass of the scattering states, and
$\Lambda$ is the cut-off currently in effect.  The pole in the
integrand is handled with the aid of the relation,
\begin{eqnarray}
\frac{1}{x+i\epsilon}&=&P\frac{1}{x}-i\pi \delta (x)\; .
\end{eqnarray}

The difference equations which determine the running of the
parameters caused by successive Bloch transformations,
without intervening rescalings, are
given by,
\begin{eqnarray}
g'_{n+1}&=&g'_{n}-\frac{3g_{n}^{'2}}{2\pi^{2}}ln\frac{\Lambda_{n}}
{\Lambda_{n+1}}-\frac{3g'_{n}w'_{n}}{2\pi^{2}}
(\frac{\Lambda_{n}^{2}-\Lambda_{n+1}
^{2}}{\Lambda_{0}^{2}}) \nonumber \\
& &-\frac{3w_{n}^{'2}}{8\pi^{2}}
(\frac{\Lambda_{n}^{4}-\Lambda_{n+1}^{4}}{
\Lambda_{0}^{4}})\;, \nonumber \\
w'_{n+1}&=&w'_{n}-\frac{9w_{n}^{'2}}{8\pi^{2}}
(\frac{\Lambda_{n}^{2}-\Lambda_{n+1}
^{2}}{\Lambda_{0}^{2}})-\frac{3g'_{n}w'_{n}}{\pi^{2}}ln\frac{\Lambda_
{n}}{\Lambda_{n+1}} \nonumber \\
 & & -\frac{3g_{n}^{'2}}{8\pi^{2}}(\frac{\Lambda_{0}^{2}}{\Lambda_
{n+1}^{2}}-\frac{\Lambda_{0}^{2}}{\Lambda_{n}^{2}})\;.
\end{eqnarray}

The results are shown in Figs. 4, 5, and 6.
Results are shown, in Fig. 5,
for the case of a marginal contribution to $u_{4}$,
and only a marginal correction to $u_{4}$ being kept. $w$ is set to
zero and is not allowed to run.  Only $g$ is allowed to run.
Also, results are shown in Fig. 6
for the case in which a marginal and the
leading irrelevant operator contributions to $u_{4}$ are kept,
and both are allowed to run.  It must be understood that the
Eqns. (20) do not govern the running parameters in these
cases, because there are no intervening rescaling operations.
Rather a slightly altered set of difference equations must be
used, Eqns. (27),
which keep track of the absolute lower cut-off.

Fig. 4 shows the cut-off dependence of an observable
when the couplings of the theory are not allowed to
run with the cut-off.  Plotted is the real part of a
two-boson to two-boson scattering amplitude as a function
of the invariant mass of the scattering states.  The interaction
is a $\phi^{4}$ interaction consisting of a marginal operator and the
leading irrelevant operator.  The marginal coupling is fixed
at .1 and the irrelevant coupling is fixed at 0.
Fig. 5
shows how this cut-off dependence is lessened when the marginal
coupling is allowed to run, and the irrelevant coupling is
left fixed at 0.  Fig. 6 shows that the cut-off dependence
is lessened even more when both couplings are allowed to run.
Deviations from cut-off independence is strongest near
the cut-off, as shown in Figs. 5 and 6.
Good results are obtained with the retention of just the
first marginal correction, for small coupling.
Better results are obtained with the retention of the first
irrelevant operator.
The irrelevant operator's effects are quite noticeable nearer
the cut-off.  Retention of all irrelevant operators should eliminate
all cut-off dependence.

Since $\frac{\vec{s}^{\:2}}{z(1-z)}$ is the invariant mass squared
of a state, the termination of the Bloch transformation after a
few orders of the interaction and after a finite number of irrelevant
operators indicates that this transformation has been expanded in
powers of the couplings and powers of the ratio $\frac{M^{2}}{
\Lambda_{0}^{2}}$.

For the case of a relevant and a marginal operator, in a two-particle
truncation, the following term must be added to the perturbing
interaction $v$,
\begin{eqnarray}
v_{m}&=&\int d\tilde{q}_{1}d\tilde{q}_{2}(\frac{m^{2}}
{q_{1}^{+}}+\frac{m^{2}}{q_{2}^{+}})|q_{1},q_{2}><q_{1},q_{2}|\; .
\end{eqnarray}
Hence, the sum of Figs. 3a, 7a, 7b, and 7c is given by,
\begin{eqnarray}
\frac{1}{2}\sum_{i}(<a|v|i><i|v|b>(\frac{1}{\epsilon_{a}-\epsilon_{i}}
+\frac{1}{\epsilon_{b}-\epsilon_{i}}))\; ,&
\end{eqnarray}
where Fig. 2c is the relevant operator vertex.

Now Fig. 7a equals,
\begin{eqnarray}
& &2
\int \frac{d\tilde{q}_{1i}d\tilde{q}_{2i}<a|q_{1i},q_{2i}>
<q_{1i},q_{2i}|b>}{\epsilon_{a\:{\rm or}\:b}-\frac{\vec{P}^{\:2}}
{P^{+}}-\frac{\vec{r}^{\:2}}{P^{+}x(1-x)}}(\frac{m^{2}}
{q_{1i}^{+}}+\frac{m^{2}}{q_{2i}^{+}})^{2} \; ,
\end{eqnarray}
gives zero when it acts upon $|a>$ or $|b>$.
Likewise Figs. 7b and 7c give zero.

So the relevant coupling, $\mu$, where $m^{2}=\mu \Lambda_{0}^{2}$,
runs only because of the rescaling, and not because of the Bloch
transformation, to second order in the transformation.

Inclusion of the one-particle sector will not change any of these
results.  Inclusion of the one- and three-particle sectors will
affect, slightly, the way the relevant coupling runs, because
of the diagram in Fig. 8.

\subsection{The Difference Equations}
In accordance with Perry \cite{3}, the rescaling step
of a renormalization group transformation is done according to
the following rules: (a) Multiply each
factor of transverse momentum that appears,
including those in the measure, by
$\frac{\Lambda_{1}}{\Lambda_{0}}$.  (b) For every field that
appears in an operator, such as four fields appearing in
the operator in Eqn. (8), multiply the operator by a
factor of $\frac{\Lambda_{0}}{\Lambda_{1}}$.  (c) Finally,
multiply the entire Hamiltonian by a factor of
$(\frac{\Lambda_{0}}{\Lambda_{1}})^{2}$.

The difference equations that are obtained with the entire
renormalization group transformation, in the two-particle truncation,
are the following,
\begin{eqnarray}
\mu_{n+1}&=&\frac{\Lambda_{0}^{2}}{\Lambda_{1}^{2}}\mu_{n}\; , \nonumber \\
g_{n+1}&=&g_{n}-\frac{3g_{n}^{2}}{2\pi^{2}}ln\frac{\Lambda_{0}}
{\Lambda_{1}}-\frac{3g_{n}w_{n}}{2\pi^{2}}(1-\frac{\Lambda_{1}
^{2}}{\Lambda_{0}^{2}}) \nonumber \\
& &-\frac{3w_{n}^{2}}{8\pi^{2}}(1-\frac{\Lambda_{1}^{4}}{
\Lambda_{0}^{4}}) \; ,\nonumber \\
w_{n+1}&=&\frac{\Lambda_{1}^{2}}{\Lambda_{0}^{2}}
w_{n}-\frac{9w_{n}^{2}}{8\pi^{2}}
\frac{\Lambda_{1}^{2}}{\Lambda_{0}^{2}}(1-\frac{\Lambda_{1}
^{2}}{\Lambda_{0}^{2}})-\frac{3g_{n}w_{n}}{\pi^{2}}
\frac{\Lambda_{1}^{2}}{\Lambda_{0}^{2}}ln\frac{\Lambda_
{0}}{\Lambda_{1}} \nonumber \\
 & & -\frac{3g_{n}^{2}}{8\pi^{2}}(1-\frac{\Lambda_{1}^{2}}{\Lambda_
{0}^{2}})\; .
\end{eqnarray}
\section{The Effective Hamiltonian}
In order to generate the effective, renormalized, Hamiltonian,
the starting cut-off must be raised toward infinity while
the couplings are sent toward their values at the fixed-point.
The number of iterations of the renormalization group transformation
it takes to get to the final lower cut-off is thereby increased
toward infinity.  With this limiting process, the running parameters
must go to a non-zero limit, or the theory is trivial.

The difference equations (30), for the
two-particle truncated scalar field, suggest that
the parameters, except the mass, do go to zero in this limit.
So the effective Hamiltonian for this
model of the scalar field, using this
fixed-point, is trivial.  However, as Fig. 9 implies, if the
couplings are small, the couplings go to zero weakly.  For example,
for a starting $g$ of .1, a decay of its value by $50\%$ occurs only
after a cut-off decrease of a factor of $10^{30}$,
or after 100 iterations of a reduction in the
cut-off by a factor of 2.  Figs. 9 and 10
also show that the irrelevant coupling does track with the
marginal coupling after some initial transience.
The irrelevant coupling depends upon the cut-off only
through a functional dependence upon the marginal coupling.
This tracking sets in long before
the marginal coupling decays appreciably.
This
means that in the renormalization process
for this model of the scalar field, the cut-off cannot
really be taken to infinity, but it can be made very large.
A calculation of a truncated model of QED is also
expected to be strictly trivial.  Assuming that with effectively
no truncation of the Fock space our methods give the correct
$\beta$-function for a theory, then our methods suggest
that non-asymptotically free theories will be strictly trivial.
In fact, our methods are expected to give
strictly non-trivial results only for asymptotically free theories.

If we expand $w$ in a power series in $g$, to second order,
in accordance with the coupling coherence of Perry and Wilson
\cite{7}, and if the difference equations (30) are used to
solve for the coefficients, we obtain,
\begin{eqnarray}
w&=&-\frac{3g^{2}}{8\pi^{2}}
\end{eqnarray}
Figs. 9 and 10 show that, after some initial transcience
caused by the intial values of $w$ and $g$, $w$ goes to this
function of $g$, for small values of $g$, after a large
number of successive renormalization group transformations.
This is also in accordance with Wilson's observation of his
general solution of the difference equations with relevant,
marginal, and irrelevant couplings \cite{8}.  That is, after
a large number of renormalization group transformations,
the final irrelevant couplings will
become strongly dependent upon the values of the
final couplings of the independent relevant and marginal
operators and only weakly dependent on the initial values
of the irrelevant couplings.

The interaction term in the Hamiltonian, after a large number
of iterations of the renormalization group transformation, then
becomes, keeping only up to the leading irrelevant operator,
{\samepage
\begin{eqnarray}
v&=&3!g\int d\tilde{q}_{1}d\tilde{q}_{2}d\tilde{q}_{3}
\frac{|q_{1},q_{2}><q_{3},q_{1}+q_{2}-q_{3}|}{q_{1}^{+}+q_{2}^{+}
-q_{3}^{+}}\cdot \\
& &\theta(\frac{\vec{P}^{\:2}
+\Lambda_{0}^{2}}{P^{+}}-\frac{\vec{q}_{1}^{\:2}}{q_{1}^{+}}-
\frac{\vec{q}_{2}^{\:2}}{q_{2}^{+}})\theta(\frac{\vec{P}^{\:2}+
\Lambda_{0}^{2}}{P^{+}}-\frac{\vec{q}_{3}^{\:2}}{q_{3}^{+}}-
\frac{(\vec{q}_{1}+\vec{q}_{2}-\vec{q}_{3})^{2}}{q_{1}^{+}+q_{2}^{+}-
q_{3}^{+}})\cdot \nonumber \\
& &(1-\frac{3g}{8\pi^{2}}(
\frac{\vec{s'}^{\:2}}{z'(1-z')\Lambda_{0}^{2}}+
\frac{\vec{t'}^{\:2}}{y'(1-y')\Lambda_{0}^{2}})) \; .\nonumber
\end{eqnarray}}
In the final Hamiltonian, all the energy-momentum variables
must be expressed in the same units.
To obtain the effective Hamiltonian, all the transverse momenta,
which are the new variables, must be multiplied by a factor
of $\frac{\Lambda_{0}}{\Lambda_{f}}$ to get the old variables.
$\Lambda_{f}$ is the final cut-off.  The entire Hamiltonian must
also be multiplied by $\frac{\Lambda_{f}^{2}}{\Lambda_{0}^{2}}$
to undo the rescaling of the energy eigenvalues of the initial
Hamiltonian.
Also, the rescaling of the fields
must be undone, so that each field variable which appears must
be multiplied by a factor of $\frac{\Lambda_{f}}{\Lambda_{0}}$.
The net result is that to obtain the effective Hamiltonian
replace every occurrence of $\Lambda_{0}$ in the final Hamiltonian
with a $\Lambda_{f}$.

Finally, the effective Hamiltonian is the following,
{\samepage
\begin{eqnarray}
{\rm H}_{eff}&=&
\int d\tilde{q}
(\frac{\vec{q}^{\:2}+\mu \Lambda_{f}^{2}}{q^{+}})
|q><q| \nonumber \\
&+&\int d\tilde{q}_{1} d\tilde{q}_{2}
\theta(\frac{\vec{P}^{\:2}
+\Lambda_{f}^{2}}{P^{+}}-\frac{\vec{q}_{1}^{\:2}}{q_{1}^{+}}-
\frac{\vec{q}_{2}^{\:2}}{q_{2}^{+}})
(\frac{\vec{q}_{1}^{\:2}+\mu \Lambda_{f}^{2}}
{q_{1}^{+}}+\frac{\vec{q}_{2}^{\:2}+\mu \Lambda_{f}^{2}}
{q_{2}^{+}})|q_{1},q_{2}><q_{1},q_{2}| \nonumber \\
 &+&3!g
\int d\tilde{q}_{1}d\tilde{q}_{2}d\tilde{q}_{3}
\frac{|q_{1},q_{2}><q_{3},q_{1}+q_{2}-q_{3}|}{q_{1}^{+}+q_{2}^{+}
-q_{3}^{+}}\cdot  \\
& &\theta(\frac{\vec{P}^{\:2}
+\Lambda_{f}^{2}}{P^{+}}-\frac{\vec{q}_{1}^{\:2}}{q_{1}^{+}}-
\frac{\vec{q}_{2}^{\:2}}{q_{2}^{+}})\theta(\frac{\vec{P}^{\:2}+
\Lambda_{f}^{2}}{P^{+}}-\frac{\vec{q}_{3}^{\:2}}{q_{3}^{+}}-
\frac{(\vec{q}_{1}+\vec{q}_{2}-\vec{q}_{3})^{2}}{q_{1}^{+}+q_{2}^{+}-
q_{3}^{+}})\cdot \nonumber \\
& &(1-\frac{3g}{8\pi^{2}}(
\frac{\vec{s'}^{\:2}}{z'(1-z')\Lambda_{f}^{2}}+
\frac{\vec{t'}^{\:2}}{y'(1-y')\Lambda_{f}^{2}})) \; .\nonumber
\end{eqnarray}}
The interaction has an extra term due to the leading irrelevant
operator.  In fact, there are an infinite number of these
extra interactions due to an infinite number of irrelevant operators,
of decreasing importance.  More of these interactions can be
systematically included.  These extra interactions represent the
effects of the high energy sector of the theory upon the low
energy sector, within this truncation of Fock space.

A straightforward working out of the second-order Bloch transformations,
within the two-particle truncation, for which the verticies, interactions,
consist of the irrelevant operators $(\frac{\vec{s}^{\:2}}{z(1-z)})^{n}$
and $(\frac{\vec{t}^{\;2}}{y(1-y)})^{m}$, and products of these,
where $n,m=1,2,...$,
do not generate marginal operators in addition to the one that we
have assumed.  That is, within the two-particle truncation, second-order
Bloch transformations with verticies that are functions of
$\frac{\vec{s}^{\:2}}{z(1-z)}$
and $\frac{\vec{t}^{\;2}}{y(1-y)}$ generate interactions that are,
in turn, functions of these quantities.  This symmetry guarantees that
no new marginal operators are generated.  The bare Hamiltonian has the
structure we have assumed for our perturbation of the fixed-point.
We do not expect this symmetry to generalize beyond the two-particle
Tamm-Dancoff truncation.  This symmetry does make the theory with a
two-particle Tamm-Dancoff truncation much easier to study than the
full theory.
\section{Conclusions}
To broaden any calculation of this type there are some obvious
and not-so-obvious options.  The order in the interaction and
the number of irrelevant operators kept, both in the renormalization
group transformation, could be increased.  The Fock space
could be increased.  It may turn out that in doing this, more
relevant and marginal operators will be generated.  One may
then implement the broader program of coupling coherence
of Perry and Wilson \cite{7,14,15}.
This program re-institutes
lost symmetries, and in the process, limits the number of
independent relevant and marginal operators.  Implementation
of the solution scheme of Wilson \cite{8} may need to
be carried out.  This method of solution of the difference
equations gives all the marginal, relevant, and irrelevant
couplings in terms of the independent relevant and marginal
couplings, given at the low cut-off.  Having obtained the
effective Hamiltonian, one then may carry out a non-perturbative,
numerical, or possibly analytical,
solution of the resulting integral
equations for the eigenvalues and wavefunctions or scattering
amplitudes.

We have carried out the procedure of Perry \cite{3}, but
in a Tamm-Dancoff truncation of Fock space.  The couplings
run differently because of the truncation.  We do not get the
same $\beta$-function as that of the nontruncated scalar field,
for example.  The Tamm-Dancoff truncation
does introduce errors, as we
expect.  However, the couplings run in the manner appropriate
for the given truncation of the Fock space.

Although the
effective Hamiltonian is trivial, which agrees with other
non-perturbative studies \cite{16}, for small coupling,
a non-trivial effective Hamiltonian can still be found.  This
non-trivial effective Hamiltonian results from making the starting
cut-off large but not infinite.  One then declares that
all higher energy scales are unimportant to the problem.
This suggests that, in this manner,
our procedure could be applied to other non-asymptotically free
theories with small couplings, such as QED.

We have shown how a non-perturbative study of QCD may be approached.
Since QCD is asymptotically free, triviality will not be an issue.
After obtaining the truncated canonical Hamiltonian,
one uses the renormalization group transformation to lower the cut-off
until the couplings are borderline perturbative.
One then implements the coupling
coherence program and solves the renormalization group difference
equations.  This gives the Hamiltonian with added effective interactions.
The rescalings are undone, and the effective, low energy Hamiltonian,
in a Tamm-Dancoff truncation, is obtained.

The Tamm-Dancoff truncation, with a good renormalization program,
is a variational approach to the field theoretic problem.  In
non-relativistic quantum mechanics the variational approach is
a powerful, popular approach.  All variational approaches come with
the caveat that certain states are described better, by a given
truncation, than other states.  All states are described equally
well with no truncation.  A Hilbert space or Fock space cannot be
put on a computer, however.  Truncation is necessary.  In this
regard, ways of generating the effective, non-local interactions
that compensate for the fact of the truncation have not been found.
Ways of finding these interactions would indeed be desirable.
Those states that are described well are well described, in part,
because of the interactions induced by removing high energies.
For these states, the missing interactions needed to compensate
for the truncation are less important.  The reverse is true for
those states that are not well described by a given Tamm-Dancoff
truncation and renormalization.  A given truncation may be horrible
for all eigenstates of the Hamiltonian.  One's knowledge must
be used, either derived from experiment or otherwise, to choose
truncations that are going to be useful.  The usefulness of one's
guesses for a truncation can be tested by looking for stability
of the results against further increases of the Fock space.

The methods presented here promise to assist in
a non-perturbative solution
of asymptotically free theories in their strong coupling regimes.
The solution of non-asymptotically free theories in their strong
coupling regimes remains an open question, from the point of
view of this approach.
\renewcommand{\thesection}{}
\section{Acknowledgements}
The theory
undergirding these calculations was introduced to me by Professors
Robert J. Perry and Kenneth G. Wilson.  Critical comments by them
and by Professors Bunny C. Clark and Richard J. Furnstahl are
greatly appreciated.
Support for this work was provided by the
Ohio State University College of
Mathematical and Physical Sciences, and by NSF grants
PHY-8858250 and PHY-9102922.

\newpage
\section{Appendix}

We want to generate the Tamm-Dancoff truncated free scalar field
Hamiltonian.  Starting with,
\begin{eqnarray}
{\rm H}_{0}&=&\int dx^{-}d^{2}x\frac{1}{2}\phi (x)(-\partial ^{\bot^{2}}
+m^{2})\phi (x)\; ,
\end{eqnarray}
which is the free Hamiltonian in the Schr\"{o}dinger representation
(it contains no light-front ``time" dependence, $x^{+}$)
where,
\begin{eqnarray}
x^{+}&=&x_{0}+x_{3}\;, \nonumber \\
x^{-}&=&x_{0}-x_{3} \;,\nonumber \\
\phi (x)&=&\phi (x^{-},x_{1},x_{2}) \nonumber \\
 &=&\int \frac{dq^{+}d^{2}q}{16\pi^{3}q^{+}}\theta (q^{+})
[a(q)e^{-iq\cdot x}+a^{\dagger}(q)e^{iq\cdot x}]\;,
\end{eqnarray}
and $(x_{0},x_{1},x_{2},x_{3})$ are normal spacetime coordinates.
\begin{eqnarray}
q^{+}&=&q_{0}+q_{3}\;, \nonumber \\
\vec{q}&=&(q_{1},q_{2})\;, \nonumber \\
a(q)&=&a(q^{+},\vec{q})\;.
\end{eqnarray}
$q^{+}$ is the `longitudinal' momentum.  $\vec{q}$ is the
`transverse' momentum.  Here the operators $a(q)$, $a^{\dagger}(q)$
satisfy,
\begin{eqnarray}
[a(q),a^{\dagger}(q')]&=&16\pi^{3}q^{+}\delta(q-q')\;,
\end{eqnarray}
and,
\begin{eqnarray}
d\tilde{q}&=&\frac{dq^{+}d^{2}q}{16\pi^{3}q^{+}}\;.
\end{eqnarray}
This implies,
\begin{eqnarray}
{\rm H}_{0}&=&\int d\tilde{q} (\frac{\vec{q}^{\:2}+m^{2}}{q^{+}})
a^{\dagger}(q)a(q)\;.
\end{eqnarray}
Now, a multi-particle ket is created with the application of
creation operators, $a^{\dagger}(q)$, where,
\begin{eqnarray}
|q_{1},q_{2},...>&=&a^{\dagger}(q_{1})a^{\dagger}(q_{2})...|0>\;.
\end{eqnarray}
The rules for the overlap of these kets can be inferred from
the following;
\begin{eqnarray}
<k|q>&=&16\pi^{3}q^{+}\delta (k^{+}-q^{+})\delta (\vec{k}-\vec{q})=
16\pi^{3}q^{+}\delta (k-q)\;,
\end{eqnarray}
\begin{eqnarray}
<k_{1},k_{2}|q_{1},q_{2}>&=&(16\pi^{3})^{2}q_{1}^{+}q_{2}^{+}
(\delta (k_{1}-q_{1})\delta (k_{2}-q_{2})+\delta (k_{1}-q_{2})
\delta (k_{2}-q_{1}))
\end{eqnarray}
$$<k_{1},k_{2},...|q_{1},q_{2},...>={\rm etc.}$$

The resolution of the identity is,
\begin{eqnarray}
1&=&|0><0|+\int d\tilde{q} |q><q|+\frac{1}{2!}\int d\tilde{q}
d\tilde{k} |q,k><q,k|+...\;,
\end{eqnarray}
and this implies,
\begin{eqnarray}
{\rm H}_{0}&=&\int d\tilde{q} (\frac{\vec{q}^{\:2}+m^{2}}{q^{+}})
|q><q|+\int d\tilde{q} d\tilde{k} (\frac{\vec{q}^{\:2}+m^{2}}
{q^{+}}+\frac{\vec{k}^{2}+m^{2}}{k^{+}})|q,k><q,k|\;+...
\nonumber \\
& &
\end{eqnarray}
If
\begin{eqnarray}
{\rm H}_{{\rm I}}&=&\frac{\lambda}{4!}\int dx^{-}d^{2}x:\phi ^{4}(x):\;,
\end{eqnarray}
then in the one-two particle truncation,
\begin{eqnarray}
{\rm H}_{{\rm I}}&=&
\frac{\lambda}{4}\int d\tilde{q}_{1}d\tilde{q}_{2}d\tilde{q}_{3}
\frac{|q_{1},q_{2}><q_{3},q_{1}+q_{2}-q_{3}|}{q_{1}^{+}+q_{2}^{+}
-q_{3}^{+}}\;.
\end{eqnarray}
\newpage
{\Large {\bf Figure Captions}}

\noindent Figure 1:  Renormalization group trajectories emanating from
a canonical surface $H_{D}$ of cutoff $\phi^{4}$ Hamiltonians.
The canonical surface intersects the critical surface C at
$\Lambda_{0}=\infty$.  $Q_{\infty}$ is the renormalized
Hamiltonian, at the cut-off $\Lambda_{f}=1$.

\vspace{.5in}

\noindent Figure 2: Diagrams describing the boson-boson interactions
of the basis set.  (a) The marginal part of the $\phi^{4}$
interaction.  (b) Leading irrelevant part of the $\phi^{4}$
interaction.  (c) Relevant part of the $\phi^{2}$ interaction.

\vspace{.5in}

\noindent Figure 3: Diagrams contributing to the second-order Bloch
transformation.

\vspace{.5in}

\noindent Figure 4: Shown is the real part of the transition amplitude, for
two (massless)-boson to two-boson scattering, versus the invariant
mass of the scattering states.  The starting cutoff is 100 units,
and it is lowered by factors of $\sqrt{2}$.  The couplings are
held fixed.  The observable is seen to change with the cutoff.

\vspace{.5in}

\noindent Figure 5: Shown is the same observable as in Fig. 4, except
that the marginal coupling, for the $\phi^{4}$ interaction,
is allowed to run according to the second-order Bloch transformations.
The cutoff dependence is dramatically reduced.

\vspace{.5in}

\noindent Figure 6: Shown is the same observable as in Figs. 4 and 5.  The
marginal and leading irrelevant couplings of the $\phi^{4}$
interaction are both allowed to run according to the second-order
Bloch transformation.  The cutoff dependence is decreased
even further.

\vspace{.5in}

\noindent Figure 7: Second-order diagrams with a combination of relevant
verticies of the $\phi^{2}$ interaction and marginal
verticies of the $\phi^{4}$ interaction.

\vspace{.5in}

\noindent Figure 8: A second-order diagram
involving the one and three-particle sectors of Fock space.

\vspace{.5in}

\noindent Figure 9: The irrelevant coupling, $w$, is plotted versus the
marginal
coupling, $g$ (solid curve).
The values of each are generated by the full
renormalization group difference equations, where the cutoff is lowered
by a factor of two at each iteration of the transformation.
The starting values, $(g_{0},w_{0})$, are $(.1,0.)$.  Approximately
100 iterations are shown.  The dashed curve shows the function
$w=-3g^{2}/(8\pi^{2})$.  After a large enough number of iterations of
the renormalization group transformation, $w$ approaches this function
of $g$.

\vspace{.5in}

\noindent Figure 10: The same quantities are plotted, as in Fig. 9, except
the starting values, $(g_{0},w_{0})$, are $(.1,.5)$, for the solid curve.
The dashed curve is the same as in Fig. 9.
\end{document}